\documentclass{jcgt}

\setciteauthor{Russell A. Brown}
\setcitetitle{Review of Three Algorithms That Build k-d Trees}

\accepted{submitted}{accepted}{published}{Editor Name}{vol}{issue}{1}{1}{year}
\seturl{http://jcgt.org/published/vol/issue/num/}

\usepackage[usenames,dvipsnames]{xcolor}
\usepackage{tcolorbox}
\usepackage{tabularx}
\usepackage{array}
\usepackage{tabto}
\usepackage{colortbl}
\tcbuselibrary{skins}

\newcolumntype{Y}{>{\raggedleft\arraybackslash}X}
\tcbset{tab1/.style={fonttitle=\bfseries\large,fontupper=\normalsize\sffamily,
colback=yellow!10!white,colframe=red!75!black,colbacktitle=Salmon!40!white,
coltitle=black,center title,freelance,frame code={
\foreach \n in {north east,north west,south east,south west}
{\path [fill=red!75!black] (interior.\n) circle (3mm); };},}}

\tcbset{tab2/.style={enhanced,fonttitle=\bfseries,fontupper=\normalsize\sffamily,
colback=white!10!white,colframe=black!50!black,colbacktitle=Salmon!40!white,
coltitle=black,center title}}

\usepackage{listings}

\usepackage{xcolor}

\definecolor{codegreen}{rgb}{0,0.6,0}
\definecolor{codegray}{rgb}{0.5,0.5,0.5}
\definecolor{codepurple}{rgb}{0.58,0,0.82}
\definecolor{backcolour}{rgb}{0.99,0.98,0.98}

\lstdefinestyle{mystyle}{
    backgroundcolor=\color{backcolour},   
    commentstyle=\color{codegreen},
    keywordstyle=\color{magenta},
    numberstyle=\tiny\color{codegray},
    stringstyle=\color{codepurple},
    basicstyle=\ttfamily\footnotesize,
    breakatwhitespace=true,         
    breaklines=true,                 
    captionpos=b,                    
    keepspaces=true,                 
    numbers=left,                    
    numbersep=5pt,                  
    showspaces=false,                
    showstringspaces=false,
    showtabs=false,                  
    tabsize=2
}

\begin{document}

\title{Review of Three Algorithms That Build \emph{k}-d Trees}

\author
       {Russell A. Brown}

\maketitle

\begin{abstract}
\small

The original description of the \emph{k}-d tree recognized that rebalancing techniques, such as used to build an AVL tree or a red-black tree, are not applicable to a \emph{k}-d tree.  Hence, in order to build a balanced \emph{k}-d tree, it is necessary to find the median of a set of data for each recursive subdivision of that set.  The sort or selection used to find the median, and the technique used to partition the set about that median, strongly influence the computational complexity of building a \emph{k}-d tree. This article describes and contrasts three \emph{k}-d tree-building algorithms that differ in their technique used to partition the set, and compares the performance of the algorithms. In addition, dual-threaded execution is proposed for one of the three algorithms.

\end{abstract}

\section{Introduction} 
\label{sec:introduction}

Bentley introduced the \emph{k}-d tree as a binary tree that stores \emph{k}-dimensional data \cite{Bentley}.  Like a standard binary tree, the \emph{k}-d tree partitions a set of data at each recursive level of the tree.  Unlike a standard binary tree that uses only one key for all levels of the tree, the \emph{k}-d tree uses $k$ keys and cycles through the keys for successive levels of the tree.  For example, to build a \emph{k}-d tree from three-dimensional points that comprise $\left(x,y,z\right)$ coordinates, the keys would be cycled as $x,y,z,x,y,z...$  for successive levels of the \emph{k}-d tree. A more elaborate scheme for cycling the keys chooses the coordinate that has the widest dispersion or largest variance to be the key for a particular level of recursion \cite{Friedman}, but this article assumes that the keys are cycled as $x,y,z,x,y,z...$ in order to simplify the exposition.

Bentley proposed that the $x$-, $y$-, and $z$-coordinates not be used as keys independently of one another, but instead that $x$, $y$, and $z$ form the most significant portions of the respective super keys $x$:$y$:$z$, $y$:$z$:$x$, and $z$:$x$:$y$ that represent cyclic permutations of $x$, $y$, and $z$.  The symbols for these super keys use a colon to designate the concatenation of the individual $x$, $y$ and $z$ values.  For example, the symbol $z$:$x$:$y$ represents a super key wherein $z$ is the most significant portion of the super key, $x$ is the middle portion of the super key, and $y$ is the least significant portion of the super key.

\newpage

\begin{figure}[h]
\centering
\centerline{\includegraphics*[trim = {1.2in, 1.5in, 1.2in, 1.5in}, clip, width=\columnwidth]{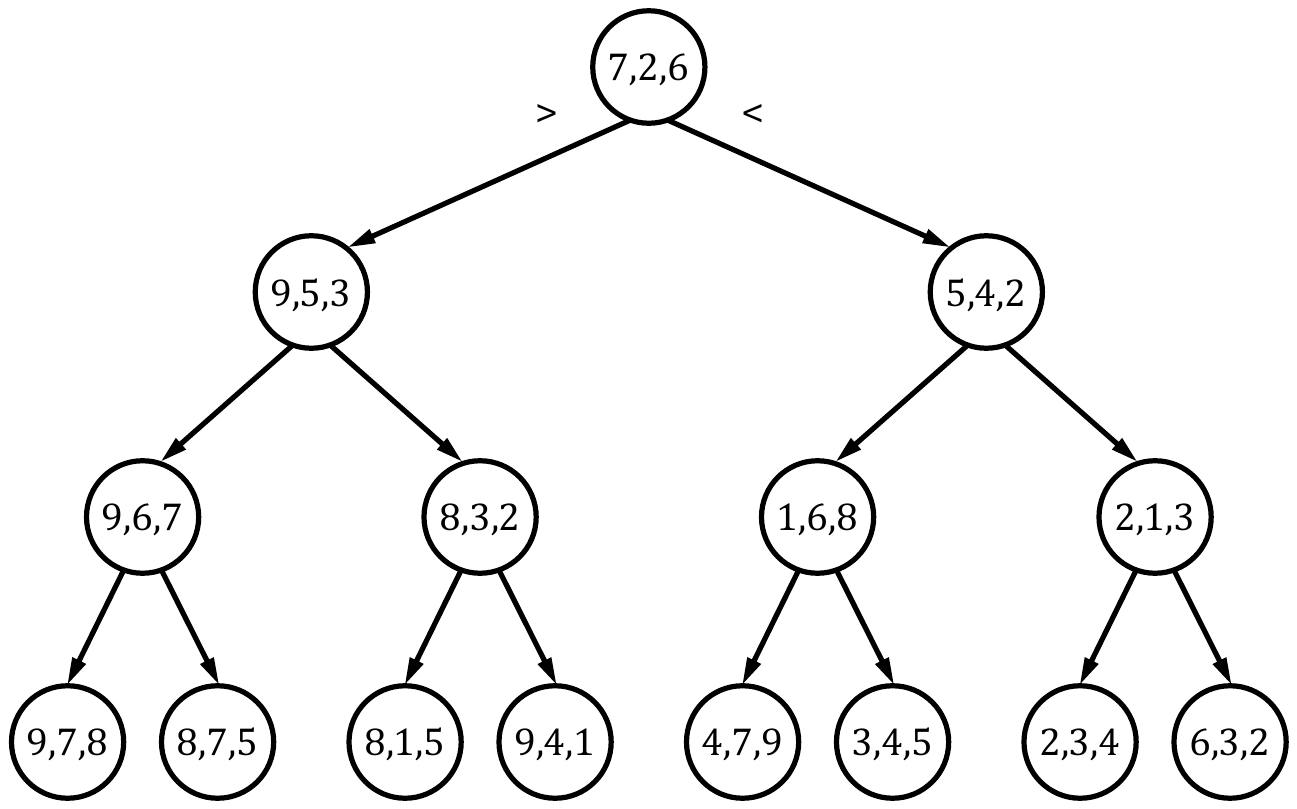}}
\caption{A \emph{k}-d tree built by recursively partitioning by $x$:$y$:$z$, $y$:$z$:$x$, and $z$:$x$:$y$ super keys}
\label{fig:kdtree}
\end{figure}

Figure \ref{fig:kdtree} depicts a balanced \emph{k}-d tree wherein 15 $\left( x,y,z \right)$ tuples are partitioned by an $x$:$y$:$z$ super key at the root of the tree, then partitioned by a $y$:$z$:$x$ super key at the first level below the root, and then partitioned by a $z$:$x$:$y$ super key at the second level below the root. The $>$ and $<$ symbols specify that each left or right child node has a larger or smaller super key respectively than its parent. This convention is chosen so that Figures \ref{fig:nlogn}, \ref{fig:knlogn}, and \ref{fig:farray} may be rotated 90 degrees clockwise to match this tree.

\section{\textbf{\textit{O}}(\textbf{\textit{n}} log \textbf{\textit{n}}) Algorithm}
\label{sec:nlogn}

\begin{figure}[h]
\centering
\centerline{\includegraphics*[trim = {1.07in, 0.3in, 1.41in, 1.52In}, clip, width=\columnwidth]{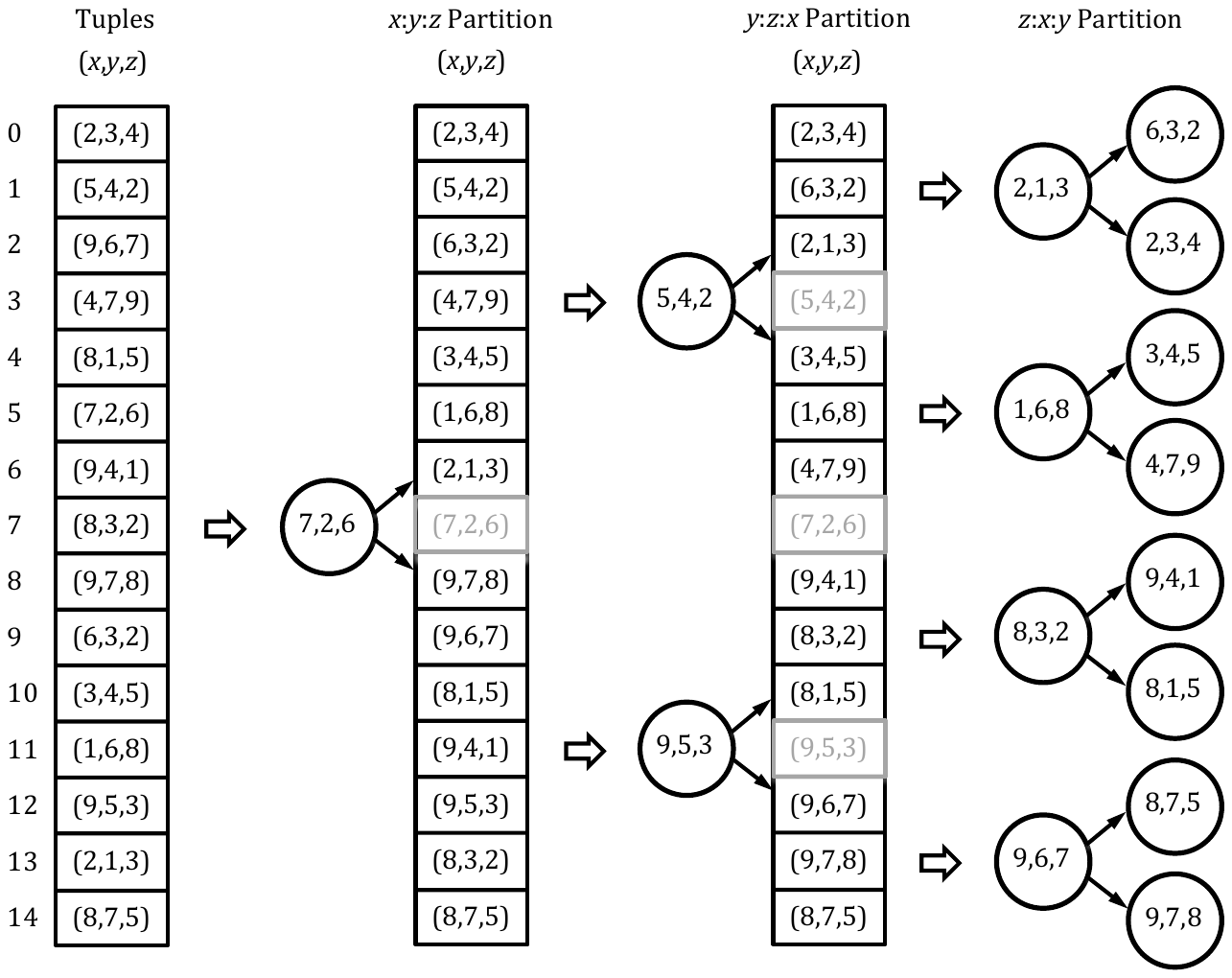}}
\caption{Building a 15-node \emph{k}-d tree via the median of medians algorithm}
\label{fig:nlogn}
\end{figure}

Figure \ref{fig:nlogn} depicts an algorithm that builds a \emph{k}-d tree in $ O\left(n \log n \right ) $ time using as input data the $\left( x,y,z \right)$ tuples shown in the Tuples array at the left side of the figure. This \emph{k}-d tree-building algorithm exploits the \emph{median of medians} algorithm proposed by Blum et al., which finds the median element of an array and partitions that array about its median element in $ O\left(n \right)$ time \cite{Blum}.

The first step of the $ O\left(n \log n \right ) $ algorithm is represented by the leftmost open arrow in Figure \ref{fig:nlogn}. This step uses the median of medians algorithm to compare $x$:$y$:$z$ super keys to find the median element $\left( 7,2,6 \right)$ of the 15-element Tuples array and to partition the Tuples array about that median element to create two 7-element sub-arrays. The median element is stored as the root node of the nascent \emph{k}-d tree, which is represented by the circle that contains ``$ 7,2,6 $" in Figures \ref{fig:kdtree} and \ref{fig:nlogn}. It remains in the Tuples array, but is ignored by further processing, as indicated by its grey color.

The $ O\left(n \log n \right ) $ algorithm proceeds recursively. Its second step is represented by the two central open arrows in Figure \ref{fig:nlogn}. This step uses the median of medians algorithm to compare $y$:$z$:$x$ super keys to find the respective median elements $\left( 5,4,2 \right)$ and $\left( 9,5,3 \right)$ of the two 7-element sub-arrays and to partition those sub-arrays about their respective median elements to create four 3-element sub-arrays. The two median elements are stored as the root node's two child nodes, which are represented by the circles that contain ``$ 5,4,2 $" and ``$9,5,3 $" in Figures \ref{fig:kdtree} and \ref{fig:nlogn}. They remain in the Tuples array, but are ignored by further processing, as indicated by their gray color.

The third step of the $ O\left(n \log n \right ) $ algorithm is represented by the four rightmost open arrows in Figure \ref{fig:nlogn}. This step uses the median of medians algorithm to compare $z$:$x$:$y$ super keys to find the respective median elements $\left( 2,1,3 \right)$, $\left( 1,6,8 \right)$, $\left( 8,3,2 \right)$, and $\left( 9,6,7 \right)$ of the four 3-element sub-arrays and to partition those sub-arrays about their respective median elements to create eight leaf nodes. The four median elements are stored as child nodes of the nodes ``$ 5,4,2 $" and ``$9,5,3 $", as per Figures \ref{fig:kdtree} and \ref{fig:nlogn}.

For the third step of the $ O\left(n \log n \right ) $ algorithm, an alternative to the median of medians algorithm partitions the 3-element sub-arrays by explicitly comparing the super keys of their three elements, which is efficient for 3-element arrays.

Two levels of recursion suffice to partition the 15-element Tuples array to create a balanced \emph{k}-d tree. If the Tuples array contained a number of elements $n$ greater than 15, further recursive partitioning would be required. In that case the median of medians algorithm would compare super keys in the order $x$:$y$:$z$, $y$:$z$:$x$, $z$:$x$:$y$... at successive levels of recursion. At each of the $\log_2 \left(n\right)$ levels of the nascent tree, the algorithm finds all of the median elements required for that level in $ O\left(n \right)$ time. Hence, the total complexity of the algorithm is $ O\left(n \log n \right ) $.

One facet of the median of medians algorithm groups the elements of an initial array into 5-element arrays, then finds the median element of each 5-element array, then groups those median elements into 5-element arrays, then finds the median element of each of those 5-element arrays, and continues to group and find until only one array remains, which contains five or fewer elements, and whose median element is the pivot point used to subdivide the initial array prior to executing that facet recursively. Due to the algorithm's heavy reliance on finding the median element, an efficient method of finding the median element is crucial to good performance.  One such method finds the median element of a 5-element array by explicitly comparing the elements hierarchically to find the median element \cite{Stepanov}.

At each level of recursion, the $ O\left(n \log n \right ) $ algorithm partitions an array into two sub-arrays with unshared and non-interleaved elements, so each sub-array may be processed by a separate thread without race conditions or excessive cache contention.

Although Figure \ref{fig:nlogn} illustrates the use of the median of medians algorithm to partition recursively a 15-element array, insertion sort finds the median of such a small array faster than the median of medians algorithm. Hence, the median of medians algorithm typically partitions a large array recursively until the sub-array size shrinks to 10-20 elements, after which insertion sort replaces the median of medians algorithm for further recursion. Also, for a sub-array that comprises two or three elements, the sub-array is partitioned by explicitly comparing the super keys of its elements.

Although, as presented above, the first step of the $ O\left(kn \log n \right ) $ algorithm uses the median of median algorithm to find the median element of the Tuples array and to partition the Tuples array about that median element, it is convenient instead to sort the Tuples array via merge sort. Then one pass through the sorted array permits removal of duplicate tuples, which is required prior to building the \emph{k}-d tree. After removal of duplicate tuples, the array is partitioned about its median element to create two sub-arrays. The second step of the $ O\left(kn \log n \right ) $ algorithm then finds the median element of each sub-array, as discussed above.

Although Figure \ref{fig:nlogn} depicts partitioning the Tuples array, a more efficient approach is to partition an array of pointers to tuples, as is the case for the $ O\left(kn \log n \right ) $ algorithm discussed subsequently in Section \ref{sec:knlogn}.

\newpage

\section{\textbf{\textit{O}}(\textbf{\textit{kn}} log \textbf{\textit{n}}) Algorithm}
\label{sec:knlogn}

\begin{figure}[h]
\centering
\centerline{\includegraphics*[trim = {1.09in, 0.29In, 1.77In, 1.50In}, clip, width=\columnwidth]{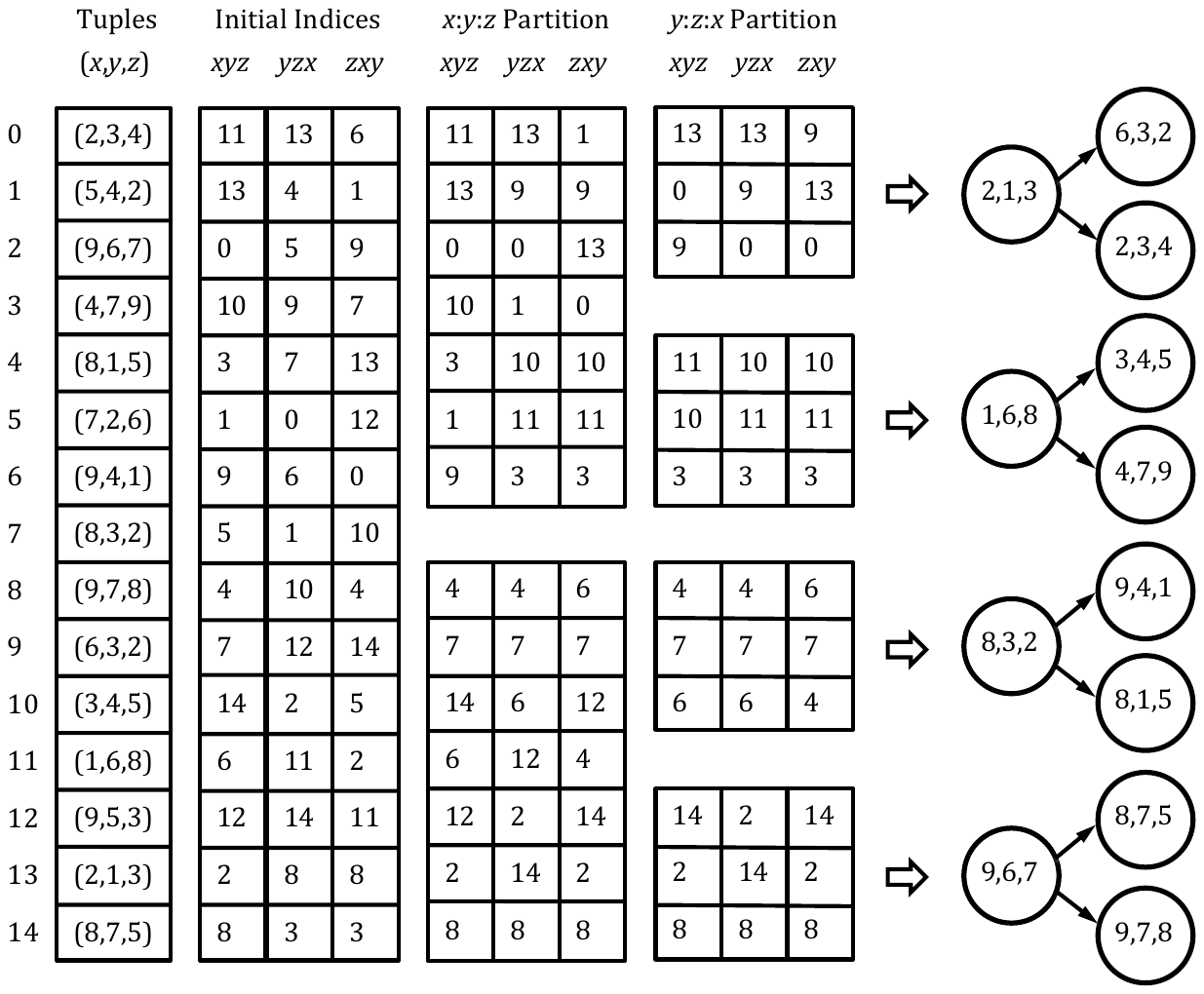}}
\caption{Building a 15-node \emph{k}-d tree after presorting tuples in each of \emph{k} dimensions}
\label{fig:knlogn}
\end{figure}

Figure \ref{fig:knlogn} depicts an algorithm that builds a \emph{k}-d tree in $ O\left(kn \log n \right ) + O\left[ \left( k-1 \right)n \log n \right] $ time using as input data the $\left( x,y,z \right)$ tuples shown in the Tuples array at the left side of the figure. This algorithm begins by presorting the tuples in each of their $x$-, $y$- and $z$-coordinates via three executions of merge sort prior to building the tree \cite{Procopiuc} \cite{Brown2015}. The $x$-, $y$-, and $z$-coordinates are not used as sorting keys independently of one another; instead, $x$, $y$ and $z$ form the most significant portions of the respective super keys $x$:$y$:$z$, $y$:$z$:$x$ and $z$:$x$:$y$ that represent cyclic permutations of $x$, $y$ and $z$.

The three merge sorts do not reorder the Tuples array; rather, they reorder three index arrays whose elements are indices into the Tuples array.  The initial order of these index arrays is established by the presorts and is shown in Figure \ref{fig:knlogn} in the $xyz$, $yzx$ and $zxy$ columns under ``Initial Indices."  In this figure, $xyz$, $yzx$ and $zxy$ are shorthand notations for the super keys $x$:$y$:$z$, $y$:$z$:$x$ and $z$:$x$:$y$ respectively. After the presorts, the $ O\left(kn \log n \right ) $ algorithm partitions the index arrays as follows.

Step 1 of the algorithm partitions the $yzx$- and $zxy$-index arrays about the median element of the $xyz$-index array using $x$:$y$:$z$ super keys. The $xyz$-index array was sorted using that key, so there is no need to partition it because it is already partitioned such that the tuples that form the smaller super keys occupy the low-address half of the $xyz$-index array that begins with the first element of the array, and the tuples that form the larger super keys occupy the high-address half of the $xyz$-index array that begins with the element just above the median element. Also, as addresses progressively increase to access elements at higher addresses within these halves of the $xyz$-index array, the indices at those addresses index tuples within the Tuples array that form progressively larger $x$:$y$:$z$ super keys. The $yzx$- and $zxy$-index arrays are partitioned in the manner explained below to adopt features similar to these features of the $xyz$-index array.

The $yzx$-index array is partitioned to create a new $yzx$-index array. The indices are retrieved from the $yzx$-index array, in order, from lowest address to highest address. For each retrieved index, an $xyz$ super key is formed from the tuple indexed by that retrieved index, and the super key is compared to the $xyz$ super key of the median element. If the retrieved index's super key is smaller than the median element's super key, the index is copied to the low-address half of the new $yzx$-index array. But if the retrieved index's super key is larger than the median element's super key, the index is copied to the high-address half of the new $yzx$-index array. And if the retrieved index's super key equals the median element's super key, the index is not copied to either half of the new $yzx$-index array, because it indexes the same tuple as the median element.

As indices are retrieved from progressively higher addresses within the $yzx$-index array, they are copied to progressively higher addresses within the low-address and high-address halves of the new $yzx$-index array. This approach partitions the $yzx$-index array such that the new $yzx$-index array adopts similar features to those of the $xyz$-index array. Specifically, the two halves of the new $yzx$-index array are partitioned about the median element of the $xyz$-index array. And as addresses increase within these halves of the new $yzx$-index array, the indices at those addresses index tuples within the Tuples array that form progressively larger $y$:$z$:$x$ super keys.

The $zyx$-index array is partitioned in an identical manner to the partitioning of the $yzx$-index array about the median element of the $xyz$-index array, such that the two halves of a new $zyx$-index array adopt similar features to those of the two halves of the new $yzx$-index array.

Step 2 of the algorithm partitions the two halves of the $xyz$-index array, and the two halves of the new $zyx$-index array, about the respective median elements of the two halves of the new $yzx$-index array using $y$:$z$:$x$ super keys. This step produces four quarters of new $xyz$- and  $zyx$-index arrays.

Step 3 of the algorithm partitions the four quarters of the new $xyz$- and $yzx$-index arrays about the respective median elements of the four quarters of the new $zyx$-index array using $z$:$y$:$x$ super keys. This step produces eight eighths of new $xyz$- and $yzx$-index arrays.

Steps 4, 5, 6, et cetera of the algorithm recursively partition the index arrays using $x$:$y$:$z$, $y$:$z$:$x$, $z$:$y$:$x$: super keys et cetera. A detailed description of the $ O\left(kn \log n \right ) $ algorithm provides further details \cite{Brown2015} .

\section{\textbf{\textit{O}}(\textbf{\textit{kn}} log \textbf{\textit{n}}) + \textbf{\textit{O}}(\textbf{\textit{n}} log \textbf{\textit{n}}) Algorithm}
\label{sec:yucao}

Cao et al. have proposed an algorithm that presorts a set of tuples in $ O\left(kn \log n \right ) $ time and then builds a \emph{k}-d tree from those tuples in $ O\left(n \log n \right ) $ time \cite{Cao}. The presorting is identical to the presorting used for the $ O\left(kn \log n \right ) $ algorithm.

This $ O\left(kn \log n \right ) + O\left(n \log n \right ) $ algorithm does not partition arrays of pointers to the Tuples array in the manner of the  $ O\left(n \log n \right ) $ and $ O\left(kn \log n \right ) $ algorithms. Instead, the algorithm registers partitioning of the Tuples array via the BN and SS registration arrays depicted in Figure \ref{fig:unsortedyucao} \cite{Cao}.

\subsection{Registration Arrays}
\label{sec:yucaoregistration}

\begin{figure}[h]
\centering
\centerline{\includegraphics*[trim = {1.07in, 0.44in, 2.57In, 1.49In}, clip, width=\columnwidth]{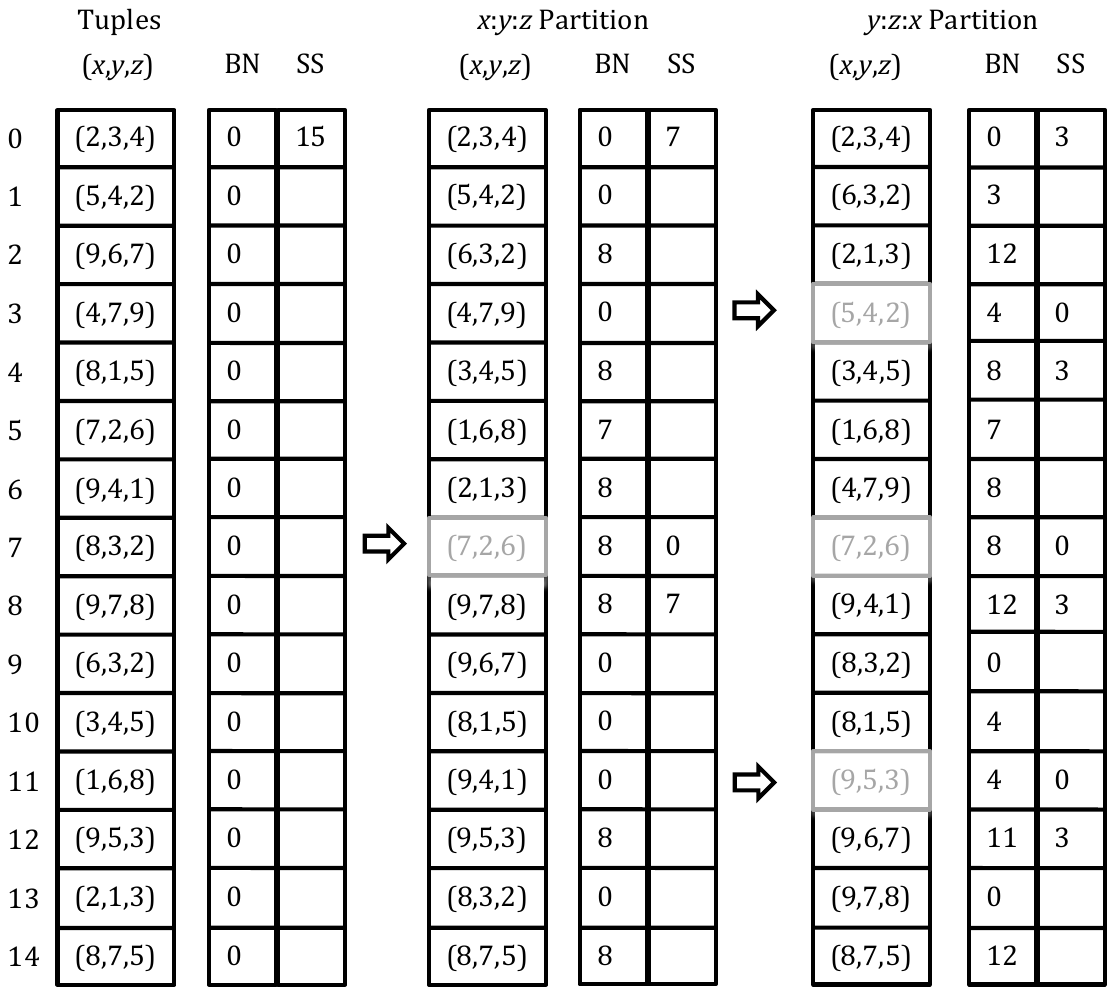}}
\caption{BN and SS array encoding of median of medians partitioning of the Tuples array}
\label{fig:unsortedyucao}
\end{figure}

Figure \ref{fig:unsortedyucao} shows how the BN and SS registration arrays would register the partitions depicted in Figure \ref{fig:nlogn} that are created by recursive partitioning of the Tuples array.

Each element $ \mathrm{BN} \negthinspace \left[ i \right] $ of the BN (aka BegiN) array contains the start index of the sub-array (within BN) to which $ \mathrm{Tuples} \negthinspace \left[ i \right] $ belongs. This start index $ \mathrm{BN} \negthinspace \left[ i \right] $ is a unique identifier for a particular sub-array. For example, the 15-element Tuples array depicted in Figure \ref{fig:unsortedyucao} contains 15 elements that all belong to a sub-array that starts at index 0, as specified by the 15 zero elements in the BN array adjacent to and immediately to the right of the Tuples array. Although the elements of the BN array index the Tuples array, a unique one-to-one mapping from an element of the BN array to an element of the Tuples array does not exist until each element of the SS array (defined below) is either 0 or 1, as depicted subsequently in Figure \ref{fig:sortedyucao}.

Each element $ \mathrm{SS} \negthinspace \left[ \mathrm{BN} \negthinspace \left[ i \right] \right] $ of the SS (aka Subtree Size) array specifies the size of a sub-array that is registered in the BN array and whose start index is $ \mathrm{BN} \negthinspace \left[ i \right] $. For example, the size of the sub-array that corresponds to all elements of the Tuples array is 15. The start index of that subtree is $ \mathrm{BN} \negthinspace \left[ 0 \right] = 0$, so its size is specified by element $\mathrm{SS} \negthinspace \left[ 0 \right] =15$ in the SS array adjacent to the BN array that contains 15 zero elements.

In Figure \ref{fig:unsortedyucao}, the contents of the BN and SS arrays register recursive partitioning of the Tuples array. Consider the two 7-element sub-arrays in the $\left(x,y,z\right)$ column under ``$x$:$y$:$z$ Partition" that are separated by the gray $\left(7,2,6\right)$ tuple. These sub-arrays are identical to the two 7-element sub-arrays created by the first partitioning step of the $ O\left(n \log n \right ) $ algorithm depicted in Figure \ref{fig:nlogn}. The BN and SS arrays immediately to the right of this $\left(x,y,z\right)$ column register that first partitioning step in the manner described below.

\newpage

Partitioning a 15-element array creates two sub-arrays: a low-address sub-array and a high-address sub-array. The size of the low-address sub-array, calculated using integer division that truncates, is $\mathrm{loSize}=15/2=7$. The size of the high-address sub-array is $\mathrm{hiSize}=\left(15-1\right)/2=7$. (The -1 in the size equation for the high-address sub-array accommodates sub-arrays that contain either an odd number of elements, e.g., 15, or an even number of elements, e.g., 14, whose high-address sub-array contains $\mathrm{hiSize}=6$ elements.)

The start index of the low-address sub-array is 0. The index of the median element equals the start index of the low-address sub-array plus the size of that sub-array, i.e., $0+7=7$. The start index of the high-address sub-array equals the index of the median element plus one, i.e., $7+1=8$. The SS array stores the size of the median element as the sentinel value 0 at the index of the median element in the SS array, so $\mathrm{SS} \negthinspace \left[ 7 \right] = 0$.

The SS array stores the sizes of the 7-element sub-arrays. This array is indexed by the start indices of these sub-arrays in the BN array. For the low-address sub-array, $\mathrm{SS} \negthinspace \left[ 0 \right] = \mathrm{loSize} = 7$. For the high-address sub-array, $\mathrm{SS} \negthinspace \left[ 8 \right] = \mathrm{hiSize} = 7$. 

The BN array stores the start indices of each element of the sub-arrays, as can be seen by inspection of the first few tuples in the Tuples array. For example, tuple $\left(2,3,4\right)$ at address 0 in the Tuples array belongs to the low-address sub-array, so $\mathrm{BN} \negthinspace \left[ 0 \right] = 0$. Similarly, tuple $\left(5,4,2\right)$ at address 1 in the Tuples array also belongs to the low-address sub-array, so $\mathrm{BN} \negthinspace \left[ 1 \right] = 0$. But tuple $\left(9,6,7\right)$ at address 2 in the Tuples array belongs to the high-address sub-array, so $\mathrm{BN} \negthinspace \left[ 2 \right] = 8$. And  tuple $\left(7,2,6\right)$ at address 5 in the Tuples array is the median element, so $\mathrm{BN} \negthinspace \left[ 5 \right] = 7$.

The ``$y$:$z$:$x$ Partition" depicted in Figure \ref{fig:unsortedyucao} partitions the two 7-element sub-arrays shown under ``$x$:$y$:$z$ Partition" to create the four 3-element sub-arrays shown under ``$y$:$z$:$x$ Partition." The BN and SS arrays are updated to register this ``$y$:$z$:$x$ Partition."

Consider the partitioning of the 7-element low-address sub-array produced by the ``$x$:$y$:$z$ Partition" that is subsequently partitioned by the ``$y$:$z$:$x$ Partition" to produce the gray median element $\left(5,4,2\right)$ and the low-address and high-address sub-arrays that contain $\mathrm{loSize}=7/2=3$ and $\mathrm{hiSize}=\left(7-1\right)/2=3$ elements respectively. 

The index of median element $\left(5,4,2\right)$ in the BN array equals the start index of the low-address 7-element sub-array plus the size of the low-address 3-element sub-array, or $0+3=3$.  The SS array stores the size of this median element as the sentinel value 0 at the index of the median element in the SS array, so $\mathrm{SS} \negthinspace \left[ 3 \right] = 0$.

The start indices of the low-address and high-address 3-element sub-arrays in the BN array are $0+0=0$ and $3+1=4$ respectively. The SS array stores the sizes of the 3-element sub-arrays; it is indexed by their start indices in the BN array. So, for the low-address sub-array, $\mathrm{SS} \negthinspace \left[ 0 \right] = \mathrm{loSize} = 3$. And for the high-address sub-array, $\mathrm{SS} \negthinspace \left[ 4 \right] = \mathrm{hiSize} = 3$.

Tuples $\left(2,3,4\right)$; $\left(6,3,2\right)$; and $\left(2,1,3\right)$ belong to the low-address 3-element sub-array. So, index 0 is registered at $\mathrm{BN} \negthinspace \left[ 0 \right]$, $\mathrm{BN} \negthinspace \left[ 9 \right]$, and $\mathrm{BN} \negthinspace \left[ 13 \right]$ because 0, 9, and 13 are the respective addresses of these tuples in the Tuples array.

Tuples $\left(3,4,5\right)$; $\left(1,6,8\right)$; and $\left(4,7,9\right)$ belong to the high-address 3-element sub-array. So, index 4 is registered at $\mathrm{BN} \negthinspace \left[ 10 \right]$, $\mathrm{BN} \negthinspace \left[ 11 \right]$, and $\mathrm{BN} \negthinspace \left[ 3 \right]$ because 10, 11, and 3 are the respective addresses of these tuples in the Tuples array..

Consider also, the partitioning of the 7-element high-address sub-array produced by the ``$x$:$y$:$z$ Partition" that is subsequently partitioned by the ``$y$:$z$:$x$ Partition" to produce the gray median element $\left(9,5,3\right)$ and the low-address and high-address sub-arrays that contain $\mathrm{loSize}=7/2=3$ and $\mathrm{hiSize}=\left(7-1\right)/2=3$ elements respectively.

The index of median element $\left(9,5,3\right)$ in the BN array equals the start index of the high-address 7-element sub-array plus the size of the low-address 3-element sub-array, or $8+3=11$. The SS array stores the size of this median element as the sentinel value 0 at the index of the median element in the SS array, so $\mathrm{SS} \negthinspace \left[ 11 \right] = 0$.

\newpage

The start indices of the low-address and high-address 3-element sub-arrays in the BN array are $8+0=8$ and $11+1=12$ respectively. The SS array stores the sizes of the 3-element sub-arrays; it is indexed by their start indices in the BN array. So, for the low-address sub-array, $\mathrm{SS} \negthinspace \left[ 8 \right] = \mathrm{loSize} = 3$. And for the high-address sub-array, $\mathrm{SS} \negthinspace \left[ 12 \right] = \mathrm{hiSize} = 3$.

Tuples $\left(9,4,1\right)$; $\left(8,3,2\right)$; and $\left(8,1,5\right)$ belong to the low-address 3-element sub-array, so index 8 is registered at $\mathrm{BN} \negthinspace \left[ 6 \right]$, $\mathrm{BN} \negthinspace \left[ 7 \right]$, and $\mathrm{BN} \negthinspace \left[ 4 \right]$ because 6, 7, and 4 are the respective addresses of these tuples in the Tuples array. 

Tuples $\left(9,6,7\right)$; $\left(9,7,8\right)$; and $\left(8,7,5\right)$ belong to the high-address 3-element sub-array, so index 12 is registered at $\mathrm{BN} \negthinspace \left[ 2 \right]$, $\mathrm{BN} \negthinspace \left[ 8 \right]$, and $\mathrm{BN} \negthinspace \left[ 14 \right]$ because 2, 8, and 14 are the respective addresses of these tuples in the Tuples array.

Figure \ref{fig:sortedyucao} shows how the BN and SS registration arrays would register the partition of the Tuples array shown in Figure \ref{fig:nlogn}. This particular use of the BN and SS arrays is a mere illustrative example to provide insight into the use of these arrays for registration. However, instead of registering partition of a Tuples array, these arrays are intended to register partition of index arrays, such as the $xyz$-, $yzx$-, and $zxy$-index arrays partitioned by the $ O\left(kn \log n \right ) $ algorithm and shown in Figure \ref{fig:knlogn}. This use of the BN and SS arrays is discussed subsequently in Section \ref{sec:yucaopartition}.

\subsection{Partitioning the Index Arrays}
\label{sec:yucaopartition}

\begin{figure}[h]
\centering
\centerline{\includegraphics*[trim = {1.07in, 0.45in, 0.40In, 1.50In}, clip, width=\columnwidth]{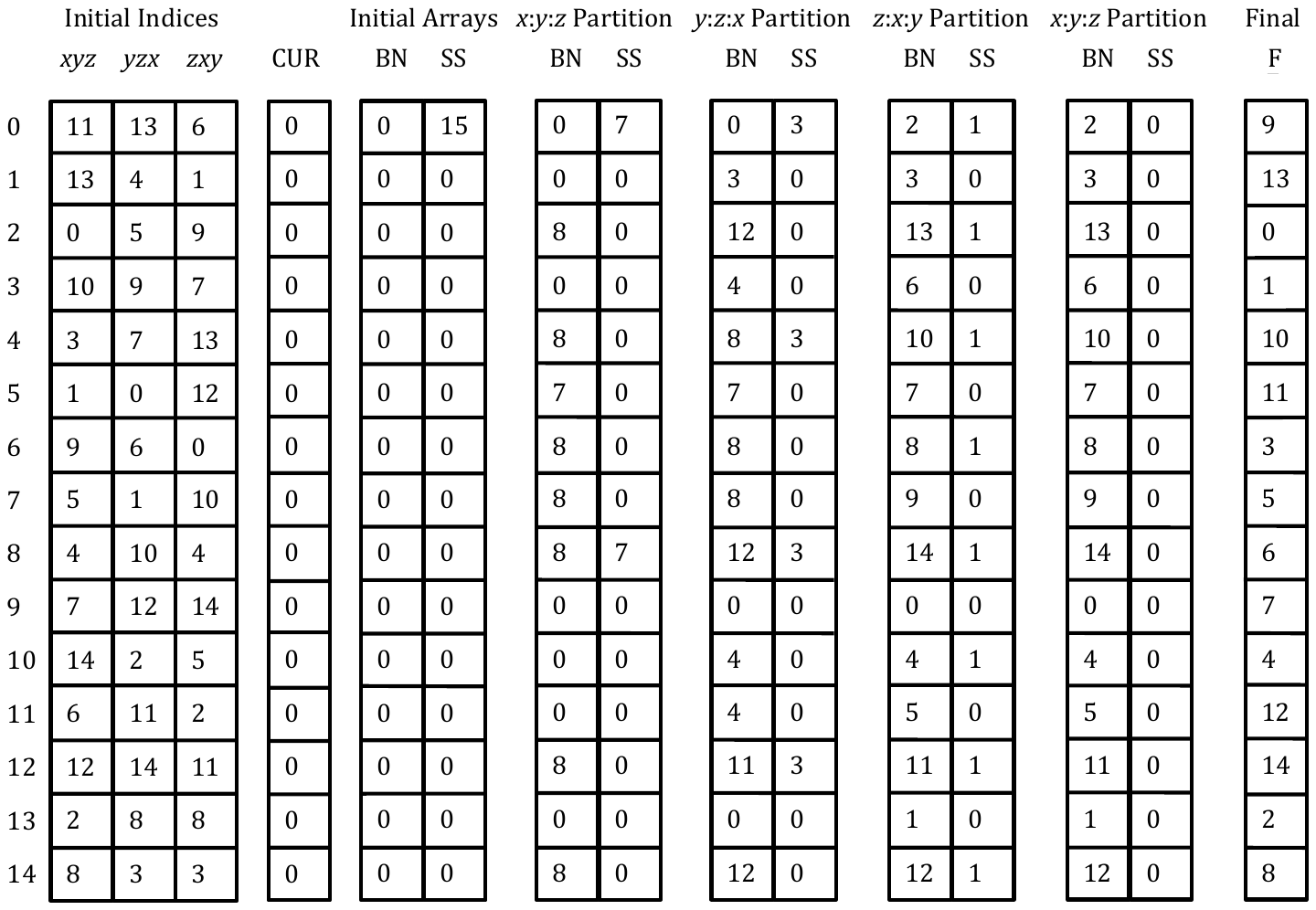}}
\caption{Partitioning the index arrays for the $ O\left(kn \log n \right ) + O\left(n \log n \right ) $  algorithm}
\label{fig:sortedyucao}
\end{figure}

Figure \ref{fig:sortedyucao} shows how the BN and SS arrays register partition of the index arrays depicted in Figure \ref{fig:knlogn}. This figure does not include the Tuples array because neither the Tuples array, nor the $x$:$y$:$z$, $y$:$z$:$x$, and $z$:$x$:$y$ super keys formed from tuples of the Tuples array, are required to register partition of the index arrays. The index arrays are not partitioned in the manner of the $ O\left(kn \log n \right ) $ algorithm that copies index-array elements. Instead, they are partitioned via registration in the BN and SS arrays.

The partitioning is facilitated by the CUR (aka CURrent element) array depicted immediately to the right of the $xyz$-, $yzx$-, and $zxy$-index arrays under ``Initial Indices" in Figure \ref{fig:sortedyucao}. Each $ \mathrm{CUR} \negthinspace \left[ \mathrm{BN} \negthinspace \left[ i \right] \right] $ element of the CUR array stores a count of the number of elements that have been retrieved from a particular index array in sorted order, and that belong to the sub-array whose start index is $\mathrm{BN} \negthinspace \left[ i \right]$ \cite{Cao}. The CUR array assists the BN and SS arrays to partition each $xyz$-, $yzx$, or $zxy$-index array into two sub-arrays, as discussed below in conjunction with Listing \ref{lst:partition}.

The C++ program of Listing \ref{lst:partition} partitions the sorted $xyz$-, $yzx$-, and $zxy$-index arrays defined by the two-dimensional \lstinline{indices} array, as per Figure 2 of \cite{Cao}. The program comprises two nested \lstinline[keywordstyle=\color{black}]{for} loops. The outer loop is initialized on line 10, cycles over the index arrays in the order $xyz$, $yzx$, $zxy$, $xyz$, ... and terminates when all elements of the SS array equal zero. The inner loop is initialized on line 12, and retrieves indices from an index array in increasing sorted order.

\newpage

For each iteration of the inner \lstinline[keywordstyle=\color{black}]{for} loop, \lstinline{tmpI} is an index retrieved from the \lstinline{indices} array in sorted order; \lstinline{tmpBN} is the start index, retrieved from the BN array, of the sub-array to which \lstinline{tmpI} currently belongs (aka the ``current" sub-array); and \lstinline{tmpSize} is the size, retrieved from the SS array, of the current sub-array. Retrieval of \lstinline{tmpI}, \lstinline{tmpBN}, and \lstinline{tmpSize} requires access to three different arrays and must be done in serial order. This requirement will be discussed further in Section \ref{sec:dualthreaded}.

Also, the \lstinline{tmpI} index permits inspection and update of \lstinline{cur[tmpI]}, which is essential for the registration of \lstinline{tmpI} via the \lstinline{bn} and \lstinline{ss} arrays, as discussed below.

The purpose of the inner \lstinline[keywordstyle=\color{black}]{for} loop is to register the \lstinline{tmpI} index by re-assigning it as the median element of the current sub-array, or by re-assigning it to either the low-address or high-address sub-array. This re-assignment is accomplished by modification of the \lstinline{bn} and \lstinline{ss} arrays. The \lstinline{cur[tmpI]} counter assists with this modification and re-assignment as follows.

First, any median element that has been registered by re-assignment in a previous iteration of the inner \lstinline[keywordstyle=\color{black}]{for} loop is ignored. Such a median element corresponds to an internal node of the nascent \emph{k}-d tree and is specified by \lstinline{tmpSize} equal to 0, as per line 16 of Listing \ref{lst:partition}.

\newpage

\begin{lstlisting}[label=lst:partition, caption=C++ program to partition the $xyz$-\, $yzx$-\, and $zxy$-index arrays]
#include <cstdlib>
int main() {
  #define DIM (3)
  #define N (15)
  size_t indices[DIM][N] = { {11,13,0,10,3,1,9,5,4,7,14,6,12,2,8},
                             {13,4,5,9,7,0,6,1,10,12,2,11,14,8,3},
                             {6,1,9,7,13,12,0,10,4,14,5,2,11,8,3} };
  size_t bn[N] = {0}, cur[N] = {0}, ss[N] = {N};
  // Loop over index arrays until all elements of SS are zero.
  for (size_t j = 0, zc = 0; zc < N; ++j) {
    size_t d = j % DIM; // Cycle index arrays as xyz, yzx, zxy ...
    for (size_t i = 0; i < N; ++i) { // Process indices in order.
      size_t tmpI = indices[d][i];
      size_t tmpBN = bn[tmpI]; // start index of current sub-array
      size_t tmpSize = ss[tmpBN]; // size of current sub-array
      if (tmpSize == 0) continue; // Skip any median element.
      size_t loSize = tmpSize/2; // size of low-address sub-array
      size_t hiSize = (tmpSize - 1)/2; // size of high-address sa
      size_t loBegin = tmpBN; // start index of low-address sa
      size_t median = tmpBN + loSize; // index of median element
      size_t hiBegin = 1 + median; // start index of hi-address sa
      if (cur[tmpBN] < loSize) bn[tmpI] = loBegin; // low-address
      else if (cur[tmpBN] > loSize) bn[tmpI] = hiBegin; // hi-addr
      else { // median element of current sub-array
        bn[tmpI] = median;
        ss[median] = 0;
        ++zc; // Increment the count of SS zero elements.
      }
      if (tmpSize > 1 && ++cur[tmpBN] >= tmpSize) { // Inc. CUR
        // Done processing current non-leaf sub-array, so split it.
        ss[loBegin] = loSize;
        ss[hiBegin] = hiSize;
        cur[loBegin] = cur[hiBegin] = 0;
      }
    }
  }
  size_t f[N];
  for (size_t i = 0; i < N; ++i) { // Create the final (F) array.
    f[bn[i]] = i;
  }
  return 0;
}
\end{lstlisting}

\newpage

Next, the count \lstinline{cur[tmpI]} is compared to the size \lstinline{loSize} of the low-address sub-array. If \lstinline{cur[tmpI]} is less than \lstinline{loSize}, then \lstinline{tmpI} is re-assigned to the low-address sub-array, as per line 22 of Listing \ref{lst:partition}, because the first \lstinline{loSize} elements counted by \lstinline{cur[tmpI]} must be partitioned into the low-address sub-array. But if \lstinline{cur[tmpI]} is greater than \lstinline{loSize}, then \lstinline{tmpI} is re-assigned to the high-address sub-array, as per line 23 of Listing \ref{lst:partition}, because the last \lstinline{hiSize} elements counted by \lstinline{cur[tmpI]} must be partitioned into the high-address sub-array. And if \lstinline{cur[tmpI]} equals \lstinline{loSize}, then \lstinline{tmpI} is re-assigned as the median element of the current sub-array, as per lines 24-28 of Listing \ref{lst:partition}.

After the above comparisons, \lstinline{cur[tmpI]} is incremented and compared to the size \lstinline{tmpSize} of the current sub-array to determine whether all indices that belong to the current sub-array have been re-assigned. If so, the current sub-array is split into its low-address and high-address sub-arrays by updating the start indices \lstinline{ss[loBegin]} and \lstinline{ss[hiBegin]}, and by initializing the counters \lstinline{cur[loBegin]} and \lstinline{cur[hiBegin]}, for those sub-arrays, as per lines 29-34 of Listing \ref{lst:partition}. But if the current sub-array comprises only one index, as specified by \lstinline{tmpSize} equal to 1, no splitting is necessary because the index \lstinline{tmpI} belongs to a leaf node of the nascent \emph{k}-d tree.

For each iteration of the inner \lstinline[keywordstyle=\color{black}]{for} loop, \lstinline{tmpSize} and \lstinline{tmpBN} permit calculation of (1) the sizes \lstinline{loSize} and \lstinline{hiSize} of the low-address and high-address sub-arrays, (2) the start indices \lstinline{loBegin} and \lstinline{hiBegin} of the low-address and high-address sub-arrays, and (3) the index \lstinline{median} of the median element of the current sub-array to which \lstinline{tmpI} currently belongs.

Upon completion of the inner \lstinline[keywordstyle=\color{black}]{for} loop, further iterations of the outer \lstinline[keywordstyle=\color{black}]{for} loop are required until all elements of the \lstinline{ss} array are equal to zero. An all-zero \lstinline{ss} array may be detected by testing each element of the \lstinline{ss} array. However, more efficient detection of an all-zero \lstinline{ss} array is implemented via the \lstinline{zc} counter that is initialized to zero in line 10, and incremented in line 27, of Listing  \ref{lst:partition}.

Figure \ref{fig:sortedyucao} depicts modification of the BN and SS arrays by the outer and inner \lstinline[keywordstyle=\color{black}]{for} loops. In this figure, the BN and SS arrays under ``Initial Arrays" define a 15-element arrays whose start index is 0. The first iteration of the outer \lstinline[keywordstyle=\color{black}]{for} loop modifies the BN and SS arrays shown under ``Initial Arrays" to produce the BN and SS arrays shown under ``$x$:$y$:$z$ Partition" by processing the $xyz$-index array under ``Initial Indices" via execution of the inner \lstinline[keywordstyle=\color{black}]{for} loop.

During execution of the inner \lstinline[keywordstyle=\color{black}]{for} loop, for each \lstinline{tmpI} index retrieved from the $xyz$-index array in order from low to high address, the corresponding element of the BN array contains 0, so \lstinline{bn[tmpI]==0} and hence \lstinline{cur[0]} counts each \lstinline{tmpI} index retrieved. Because $\mathrm{SS} \left[ 0 \right] == 15 $, \lstinline{tmpSize==15} and \lstinline{loSize==7}. Hence, the first 7 indices retrieved from the $xyz$-index array are registered into a low-address 7-element sub-array whose start index \lstinline{loBegin==0}, the 8th index retrieved from the $xyz$-index array is registered as the median element whose index is \lstinline{median==7}, and the last 7 indices retrieved from the $xyz$-index array are registered into a high-address 7-element sub-array whose start index is \lstinline{hiBegin==8}. These registrations may be verified by inspection of the BN and SS arrays under ``$x$:$y$:$z$ Partition." Specifically, the elements of the BN array at addresses 11, 13, 0, 10, 3, 1, and 9 all contain 0; $ \mathrm{BN} \left[ 5 \right] == 7$; and the elements of the BN array at addresses 4, 7, 14, 6, 12, 2, and 8 all contain 8.

At this point, the BN and SS arrays under ``$x$:$y$:$z$ Partition" define two 7-element arrays whose start indices are 0 and 8.

The second iteration of the outer \lstinline[keywordstyle=\color{black}]{for} loop modifies the BN and SS arrays shown under ``$x$:$y$:$z$ Partition" to produce the BN and SS arrays shown under ``$y$:$z$:$x$ Partition" by processing the $yzx$-index array under ``Initial Indices" via execution of the inner \lstinline[keywordstyle=\color{black}]{for} loop. The modified BN and SS arrays under ``$y$:$z$:$x$ Partition" define four 3-element sub-arrays. Two of these 3-element sub-arrays are created by splitting the 7-element array whose start index is 0; their start indices are 0 and 4. The other two of these 3-element sub-arrays are created by splitting the 7-element array whose start index is 8; their start indices are 8 and 12. For these four 3-element sub-arrays, the inner \lstinline[keywordstyle=\color{black}]{for} loop will assign \lstinline{loSize=3} and \lstinline{hiSize=3}.

During execution of the inner \lstinline[keywordstyle=\color{black}]{for} loop, the first \lstinline{tmpI} index retrieved from the $yzx$-index array under ``Initial Indices" is 13. In the BN array under ``$x$:$y$:$z$ Partition," $ \mathrm{BN} \left[ 13 \right] == 0 $ and $ \mathrm{SS} \left[ 0 \right] == 7 \ne 0 $, so \lstinline{cur[0]} counts index 13 for registration into either the 3-element sub-array whose start index is 0, or the 3-element sub-array whose start index is 4. Because index 13 is the first index \lstinline{cur[0]} counted by \lstinline{cur[0]}, \lstinline{cur[0]==0}, which is less than \lstinline{loSize==3}, so index 13 is registered into the 3-element sub-array whose start index is 0, as verified by $ \mathrm{BN} \left[ 13 \right] == 0 $ under ``$y$:$z$:$x$ Partition."

During execution of the inner \lstinline[keywordstyle=\color{black}]{for} loop, the second \lstinline{tmpI} index retrieved from the $yzx$-index array under ``Initial Indices" is 4. In the BN array under ``$x$:$y$:$z$ Partition," $ \mathrm{BN} \left[ 4 \right] == 8 $ and $ \mathrm{SS} \left[ 8 \right] == 7  \ne 0 $, so \lstinline{cur[8]} counts index 4 for registration into either the 3-element sub-array whose start index is 8, or the 3-element sub-array whose start index is 12. Because index 4 is the first index that \lstinline{cur[8]} counts, \lstinline{cur[8]==0}, which is less than \lstinline{loSize==3}, so index 4 is registered into the 3-element sub-array whose start index is 8, as verified by $ \mathrm{BN} \left[ 4 \right] == 8 $ under ``$y$:$z$:$x$ Partition."

During execution of the inner \lstinline[keywordstyle=\color{black}]{for} loop, the third \lstinline{tmpI} index retrieved from the $yzx$-index array under ``Initial Indices" is 5. In the BN array under ``$x$:$y$:$z$ Partition," $ \mathrm{BN} \left[ 5 \right] == 7 $ and $ \mathrm{SS} \left[ 7 \right] == 0 $, so index 5 is the median element of the original 15-element array and hence index 5 is not processed further by the inner \lstinline[keywordstyle=\color{black}]{for} loop.

During execution of the inner \lstinline[keywordstyle=\color{black}]{for} loop, the penultimate \lstinline{tmpI} index retrieved from the $yzx$-index array under ``Initial Indices" is 8. In the BN array under ``$x$:$y$:$z$ Partition," $ \mathrm{BN} \left[ 8 \right] == 8 $ and $ \mathrm{SS} \left[ 8 \right] == 7 \ne 0 $, so \lstinline{cur[8]} counts index 8 for registration into either the 3-element sub-array whose start index is 8, or the 3-element sub-array whose start index is 12. Because index 8 is the last index that \lstinline{cur[8]} counts, \lstinline{cur[8]==6}, which is greater than \lstinline{loSize==3}, so index 8 is registered into the 3-element sub-array whose start index is 12, as verified by $ \mathrm{BN} \left[ 8 \right] == 12 $ under ``$y$:$z$:$x$ Partition."

During execution of the inner \lstinline[keywordstyle=\color{black}]{for} loop, the last \lstinline{tmpI} index retrieved from the $yzx$-index array under ``Initial Indices" is 3. In the BN array under ``$x$:$y$:$z$ Partition," $ \mathrm{BN} \left[ 3 \right] == 0 $ and $ \mathrm{SS} \left[ 0 \right] == 7 \ne 0 $, so \lstinline{cur[0]} counts index 3 for registration into either the 3-element sub-array whose start index is 0, or the 3-element sub-array whose start index is 4. Because index 3 is the last index that \lstinline{cur[0]} counts, \lstinline{cur[0]==6}, which is less than \lstinline{loSize==3}, so index 3 is registered into the 3-element sub-array whose start index is 4, as verified by $ \mathrm{BN} \left[ 3 \right] == 4 $ under ``$y$:$z$:$x$ Partition."

At this point, the BN and SS arrays under ``$y$:$z$:$x$ Partition" define four 3-element arrays whose start indices are 0, 4, 8, and 12.

The third iteration of the outer \lstinline[keywordstyle=\color{black}]{for} loop modifies the BN and SS arrays shown under ``$y$:$z$:$x$ Partition" to produce the BN and SS arrays shown under ``$z$:$x$:$y$ Partition" by processing the $zxy$-index array under ``Initial Indices" via execution of the inner \lstinline[keywordstyle=\color{black}]{for} loop. The modified BN and SS arrays under ``$z$:$x$:$y$ Partition" have the following features. Each element of the SS array is either 0 or 1, so a unique one-to-one mapping from each element of the BN array to an element of the Tuples array depicted in Figure \ref{fig:unsortedyucao} exists. The SS array contains seven elements equal to 0, which define median elements that correspond to internal nodes of the nascent \emph{k}-d tree. The SS array contains eight elements equal to 1, which define 1-element sub-arrays.

The fourth iteration of the outer \lstinline[keywordstyle=\color{black}]{for} loop transforms these eight 1-element sub-arrays into eight median elements, because the single element of a 1-element array is by definition its median element. Given that each element of the 15-element SS array now equals 0, as determined by \lstinline{zc==15}, execution of the outer \lstinline[keywordstyle=\color{black}]{for} loop terminates.

Four iterations of the outer \lstinline[keywordstyle=\color{black}]{for} loop are required to partition the $xyz$-, $yzx$-, and $zxy$-index arrays to the point that each element of the 15-element BN array corresponds to an element of the Tuples array depicted in Figure \ref{fig:unsortedyucao}. Each iteration of the inner \lstinline[keywordstyle=\color{black}]{for} requires retrieval of 15 indices from one of the index arrays. So, the computational complexity of partitioning is $4 \times 15 \approx \log_2 \left( 15 \right) \times 15 $, or $ O \left( n \log n \right) $.

\subsection{Building the \emph{k}-d Tree}
\label{sec:yucaobuild}

\begin{figure}[h]
\centering
\centerline{\includegraphics*[trim = {1.08in, 0.32in, 1.39In, 1.50In}, clip, width=\columnwidth]{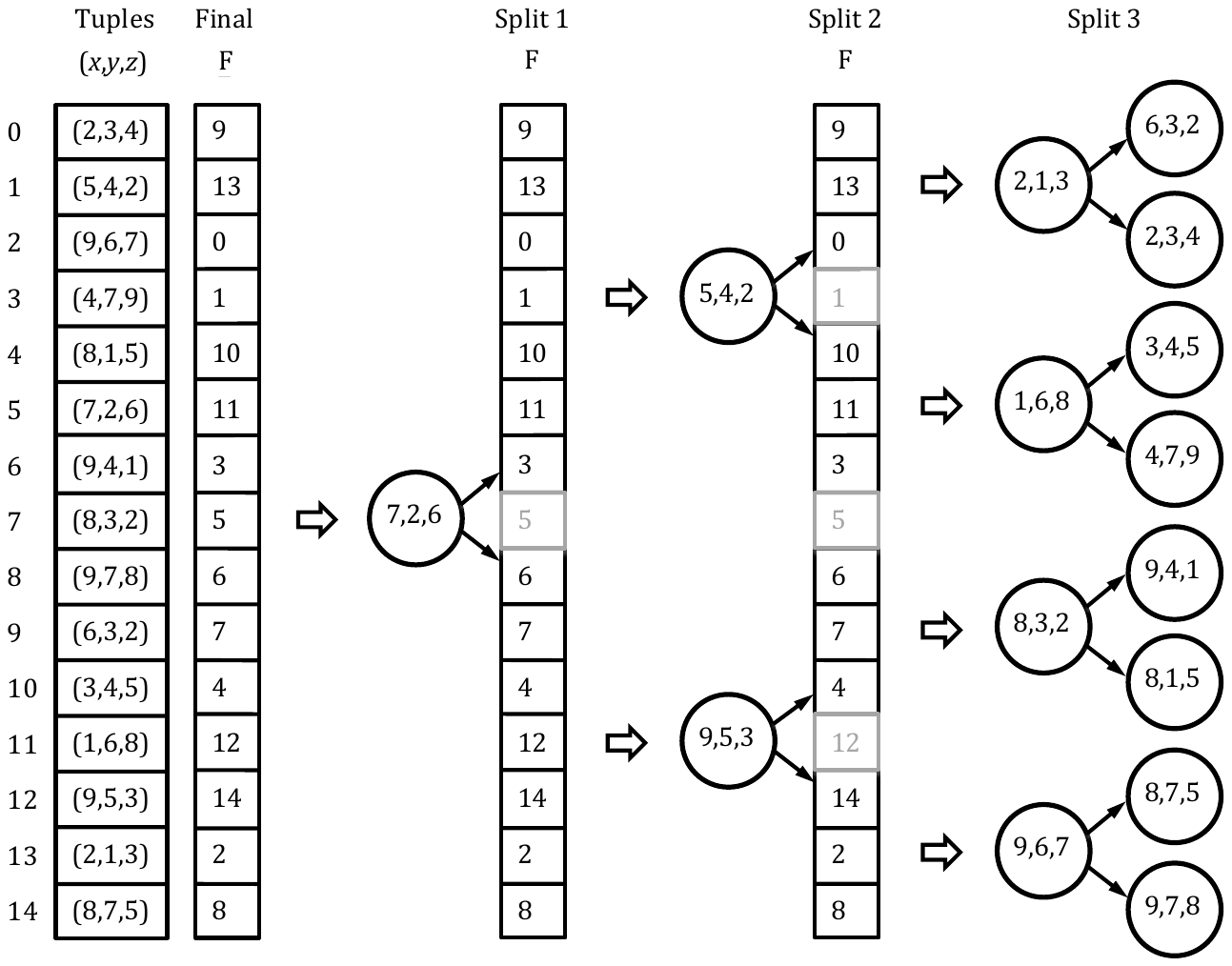}}
\caption{Building a 15-node \emph{k}-d tree from the final (F) array}
\label{fig:farray}
\end{figure}

Given the unique one-to-one mapping from each element of the BN array to an element of the Tuples array, either the BN array under ``$z$:$x$:$y$ Partition" or the BN array under the rightmost ``$x$:$y$:$z$ Partition" of Figure \ref{fig:sortedyucao} may be used to build the \emph{k}-d tree. Building the \emph{k}-d tree via a recursive algorithm is facilitated by reordering the elements (aka indices) of the BN array so that the reordered elements will index the Tuples array in sorted order \cite{Cao}. This reordering is accomplished by lines 37-40 of Listing \ref{lst:partition} that create the F (aka Final) array depicted in Figures \ref{fig:sortedyucao} and \ref{fig:farray}, wherein $ \mathrm{F} \left[ \mathrm{BN} \left[ i \right] \right] = i $ for the address $i$ of each element of the BN array.

\newpage

Figure \ref{fig:farray} depicts building the \emph{k}-d tree in $ O \left( n \right) $ time by recursively subdividing the F array in the manner prescribed in Figure 3 of the article by Yu Cao et al. \cite{Cao}. 

Subdivision begins with a 15-element F array whose start index \lstinline{tmpBN==0} and whose size \lstinline{tmpSize==15}. Assignment to the variables \lstinline{loSize=7}, \lstinline{hiSize=7}, \lstinline{loBegin=0}, \lstinline{median=7}, and \lstinline{hiBegin=8} occurs similar to lines 17-21 of Listing \ref{lst:partition}. The F array is split, as shown by the leftmost open arrow in Figure \ref{fig:farray}, to produce two 7-element sub-arrays and the root node of the nascent \emph{k}-d tree, whose tuple $\left(7,2,6\right)$ is retrieved from address 5 of the Tuples array, as indicated by the gray index 5 at address \lstinline{median==7} of the F array. The sub-arrays have respective start indices \lstinline{loBegin==0} and \lstinline{hiBegin==8}, and sizes \lstinline{loSize==7} and \lstinline{hiSize==7}.

Each 7-element sub-array is processed at the next level of recursion, which is represented by the two central open arrows in Figure \ref{fig:farray}. This level of recursion creates four 3-element sub-arrays and two internal nodes, whose tuples $\left(5,4,2\right)$ and $\left(9,5,3\right)$ are retrieved from addresses 1 and 12 of the Tuples array, as indicated by the gray indices 1 and 12 at addresses 3 and 11 of the F array respectively.

\newpage

At the final level of recursion, the four rightmost open arrows represent the creation of 12 \emph{k}-d tree nodes from the four 3-element sub-arrays, identical to the creation of 12 \emph{k}-d tree nodes by the $ O \left( kn \log n \right) $ algorithm depicted in Figure \ref{fig:knlogn}.

Although Figure \ref{fig:farray} depicts creation of \emph{k}-d nodes from 3-element sub-arrays at the final level of recursion, \emph{k}-d nodes may be created from 2-, 3-, and 4-element sub-arrays at this level of recursion.

\subsection{Dual-threaded Execution}
\label{sec:dual}

Cao et al. report that the $ O \left( n \log n \right) $ \emph{k}-d tree-building phase of the $ O \left( kn \log n \right) + O \left( n \log n \right) $ algorithm does not employ parallelism \cite{Cao}. However, dual-threaded execution of this phase is possible in a manner similar to that of the $ O \left( n \log n \right) $ and $ O \left( kn \log n \right) $ algorithms. The $ O \left( n \log n \right) $ algorithm calculates medians via two threads. The $ O \left( kn \log n \right) $ algorithm copies between index arrays for both merge sorting and index-array partitioning via two threads \cite{Brown2025}.

The $ O \left( n \log n \right) $  \emph{k}-d tree-building phase of the $ O \left( kn \log n \right) + O \left( n \log n \right) $ algorithm may similarly execute via two threads. The inner \lstinline[keywordstyle=\color{black}]{for} loop of Listing \ref{lst:partition} partitions an index array by processing array elements in increasing sorted order from the lowest address through the highest address of the index array. It is equally possible to partition an index array by processing array elements in decreasing sorted order from the highest address through the lowest address. And it is possible to partition an index array by processing array elements in increasing sorted order from the lowest address to the median address via one thread, while simultaneously processing array elements in decreasing sorted order from the highest address through the median address via a second thread.

Dual-threaded execution requires two instances of the inner \lstinline[keywordstyle=\color{black}]{for} loop of Listing \ref{lst:partition}, so that a unique inner \lstinline[keywordstyle=\color{black}]{for} loop may be executed independently by each thread. Each thread must maintain its own \lstinline{cur} array. Each thread must be able to read from, but not write to, the other thread's \lstinline{cur} array, so there is no need for these \lstinline{cur} arrays to be atomic. Although the \lstinline{bn} and \lstinline{ss} arrays are shared by the two instances of the inner \lstinline[keywordstyle=\color{black}]{for} loop, the two threads write to different elements of the \lstinline{bn} array and only read from the \lstinline{ss} array, so there is no need for these arrays to be atomic. The \lstinline{zc} counter must be atomic to permit either thread to increment it.

\section{Memory Requirements}
\label{sec:memory}

Each \emph{k}-d tree-building algorithm requires arrays in addition to the Tuples array. The $ O \left( n \log n \right) $ algorithm requires an array of pointers to the elements of the Tuples array, as well as a second array of pointers to be used as temporary storage by merge sort and by the median of medians algorithm. The $ O \left( kn \log n \right) $ algorithm requires \emph{k} arrays of pointers to the elements of the Tuples array, as well as one additional array of pointers to be used as temporary storage by merge sort and by index-array partitioning. The $ O \left( kn \log n \right) + O \left( n \log n \right) $ algorithm requires \emph{k} index arrays, as well as one additional index array for merge sort, plus the three BN, SS, and CUR arrays.

\section{Benchmarks}
\label{sec:benchmarks}

\subsection{Benchmark Methodology}
\label{sec:methodology}

To compare the performance of the three variants of the \emph{k}-d tree, benchmarks were executed on a Hewlett-Packard Pro Mini 400 G9 with 2x32GB DDR5-4800 RAM and a 14th-generation Intel Raptor Lake CPU (i7 14700T with 8 performance cores that each support 2 hyper-threads, 5.2GHz performance core maximum frequency, 78.6GB/s maximum memory bandwidth, 80KB L1 and 2MB L2 per-core caches, and a 33MB L3 cache shared by all cores).

Benchmarks for each \emph{k}-d tree variant were executed for trees that comprise $n$ nodes, where $n$ is an integer $i$ power of 2 for $i$ in the range $ \left[ 2^{16}, 2^{24} \right] $ and where each node of the tree stores a $k$-dimensional tuple of 64-bit integers. The integers were equally spaced across the maximum 64-bit integer range $r$, so the spacing was $r/n$. The integers were randomly shuffled via the \lstinline{std::mt19937_64} Mersenne Twister pseudo-random number generator \cite{Matsumoto} and copied to the first of the $k$ dimensions, then randomly shuffled again and copied to the second of the $k$ dimensions, etc.

So that all benchmarks randomly shuffled the integers in an identical sequence, each benchmark initialized \lstinline{std::mt19937_64} to \lstinline{std::mt19937_64::default_seed}.

A benchmark for each \emph{k}-d tree variant was implemented in C++, compiled via Gnu g++ 13.2.0 with the \lstinline{-std=c++11}, \lstinline{-O3}, \lstinline{-pthread}, and \lstinline{-D PREALLOCATE} options, and executed under Ubuntu 24.04.1 LTS via 1 to 16 threads mapped to up to 8 performance cores (2 threads per core) specified via the Ubuntu \lstinline{taskset} command.

Each benchmark measured the execution times for merge sort, \emph{k}-d tree construction, array allocation and deallocation, \emph{k}-d tree verification for correct ordering, etc. via the \lstinline{std::chrono::steady_clock::now()} function. Each benchmark was repeated 10 times and the mean values and standard deviations of the execution times were calculated. For the merge sort and \emph{k}-d tree construction times, which dominate the total execution time, the standard deviations were less than 5\% of the mean values.

\subsection{Dual-threaded Benchmarks}
\label{sec:dualthreaded}

Table \ref{table:executiontimes} reports the execution times in seconds for single-threaded and dual-threaded execution of the $ O \left( n \log n \right) $ \emph{k}-d tree-building phase of the $ O \left( kn \log n \right) + O \left( n \log n \right) $ algorithm for \emph{k}-d trees that comprise between $ 2^{16} $ and $ 2^{24} $ nodes, wherein each node stores a 3-dimensional tuple. This table reveals that the execution times for single-threaded and dual-threaded execution of the $ O \left( n \log n \right) $ \emph{k}-d tree-building phase are equal, with the exception of minor differences for $ 2^{16} $, $2^{22} $, and $ 2^{23} $ tuples.

\newcolumntype{T}{>{\hsize=1.5\hsize}X}

\begin{table}[htb]

\begin{tcolorbox}[tab2,tabularx={T||Y|Y|Y|Y|Y|Y|Y|Y|YYYY}]
Threads &  $2^{16}$ & $2^{17}$ & $2^{18}$ & $2^{19}$ & $2^{20}$ & $2^{21}$ & $2^{22}$ & $2^{23}$ & $2^{24}$  \\\hline\hline
1 & $6.3 \times 10^{-3}$ &  $1.4 \times 10^{-2}$ & $3.3 \times 10^{-2}$ & $7.8 \times 10^{-2}$ & $2.0 \times 10^{-1}$ &
   $5.0 \times 10^{-1}$ & $1.6 \times 10^{0}$ & $4.7 \times 10^{0}$ & $1.2 \times 10^{1}$  \\\hline
2 & $6.2 \times 10^{-3}$ & $1.4 \times 10^{-2}$ & $3.3 \times 10^{-2}$ & $7.8 \times 10^{-2}$ & $2.0 \times 10^{-1}$ &
   $5.0 \times 10^{-1}$ & $1.7 \times 10^{0}$ & $5.1 \times 10^{0}$  & $1.2 \times 10^{1}$  \\\hline
\end{tcolorbox}

\caption{\label{table:executiontimes}
Execution times (s) for the $ O\left(n \log n \right ) $ \emph{k}-d tree-building phase}

\end{table}

\begin{figure}[h]
\centering
\centerline{\includegraphics*[trim = {1.00in, 3.47in, 1.37In, 1.52In}, clip, width=\columnwidth]{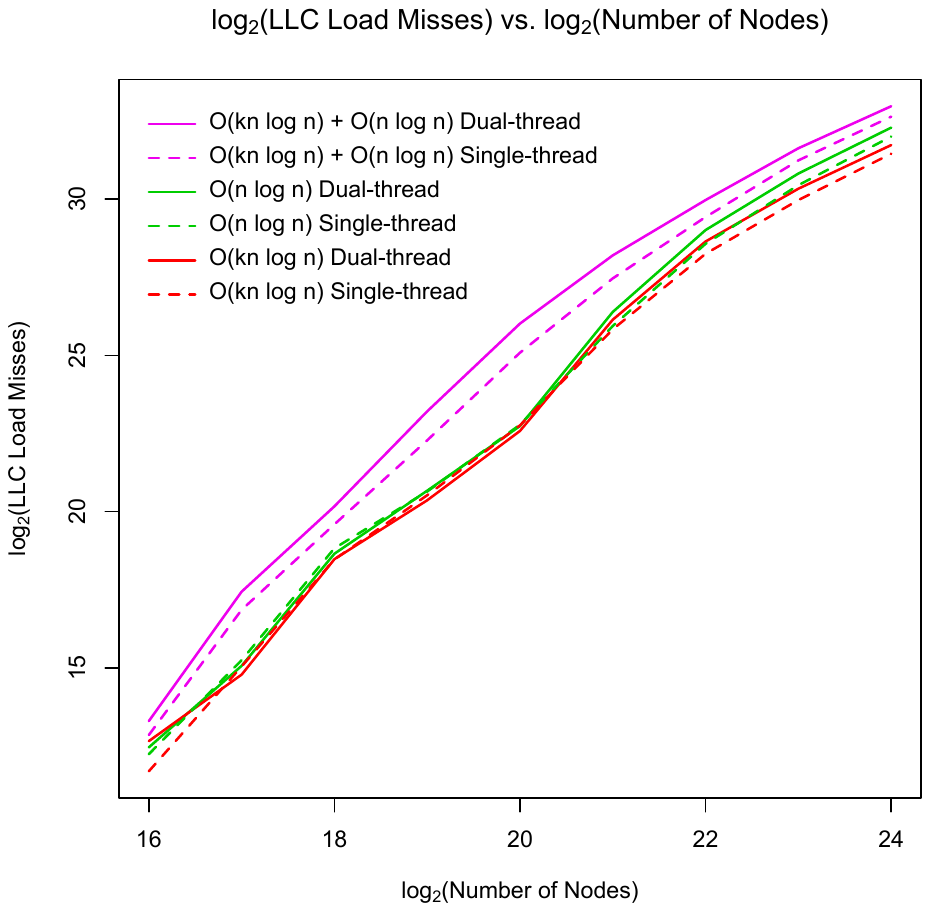}}
\caption{LLC-load-misses for three \emph{k}-d tree-building algorithms}
\label{fig:llc}
\end{figure}

Figure \ref{fig:llc} provides insight into the fact that dual-threaded execution is no faster than single-threaded execution. This figure shows log-log plots of the cache memory LLC load misses, measured via the Ubuntu \lstinline{perf stat -e LLC-load-misses} command, that occur for execution of the $ O \left( n \log n \right) $, $ O \left( kn \log n \right) $, and $ O \left( kn \log n \right) + O \left( n \log n \right) $ algorithms to construct a \emph{k}-d tree, plotted versus the number of nodes in the tree, for trees that comprise between $ 2^{16} $ and $ 2^{24} \left(x,y,z\right) $ tuples.

In this figure, the solid and dashed magenta curves for respective dual- and single-threaded execution of the $ O \left( kn \log n \right) + O \left( n \log n \right) $ algorithm are separated by one $ \log_2 $ unit in the $y$-coordinate. This $y$-separation in log space corresponds to multiplication in linear space, so dual-threaded execution exhibits twice as many LLC load misses as single-threaded execution. Similarly, the green and red curves reveal that dual- or single-threaded execution of the $ O \left( n \log n \right) $ or $ O \left( kn \log n \right) $ algorithm exhibits far fewer LLC load misses than the $ O \left( kn \log n \right) + O \left( n \log n \right) $ algorithm.

Listing \ref{lst:partition} provides an explanation for the increased number of LLC load misses for dual-threaded execution relative to single-threaded execution. For dual-threaded execution, one thread reads from the \lstinline{indices} array in ascending sequential order from the array's lowest address to its median address, while the other thread reads from the \lstinline{indices} array in descending sequential order from the array's highest address to its median address. The two threads would not be expected to experience cache conflicts when reading the \lstinline{indices} array, because each thread reads from an unshared and non-interleaved sub-array of the \lstinline{indices} array.

However, the two threads also read from the \lstinline{bn} and \lstinline{ss} arrays, with no guarantee that the threads will read from non-interleaved elements of the \lstinline{bn} and \lstinline{ss} arrays. When two threads read interleaved elements, each thread invalidates cache lines that were loaded by the other thread. This cache-line invalidation is known as \emph{false sharing} in the OpenMP literature. It occurs when multiple threads access elements of an array in an interleaved manner, which impedes OpenMP execution \cite{Chandra}. Similar cache-line invalidation may occur for dual-threaded execution of the $ O \left( n \log n \right) $ \emph{k}-d tree-building phase of the $ O \left( kn \log n \right) + O \left( n \log n \right) $ algorithm.

Moreover, when two threads access the \lstinline{indices} array, the \lstinline{tmpI} index that is read from the \lstinline{indices} array is then used to read from the \lstinline{bn} array, and then the \lstinline{tmpBN} index that is read from the \lstinline{bn} array is used to read from the \lstinline{ss} array. During these sequential reads from the arrays, a read from one array can invalidate a cache line that was loaded for another array. This cache-line invalidation may occur more frequently for dual-threaded execution than for single-threaded execution, and impede execution.

\subsection{Multi-threaded Benchmarks}
\label{sec:multihreaded}

\begin{figure}[h]
\centering
\centerline{\includegraphics*[trim = {1.00in, 3.47in, 1.37In, 1.52In}, clip, width=\columnwidth]{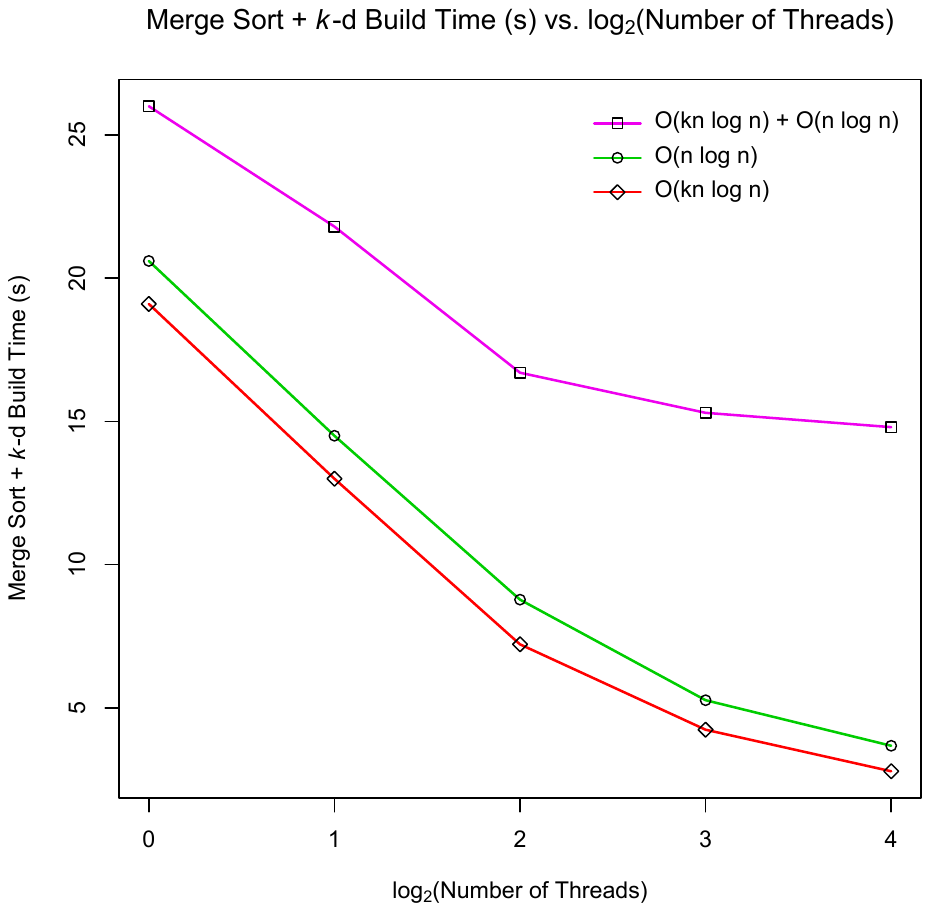}}
\caption{Execution time (s) vs. $ \log_2 $(number of threads) for $ 2^{24} \left(x,y,z\right) $ tuples}
\label{fig:time}
\end{figure}

Figure \ref{fig:time} shows semi-log plots of the times in seconds for execution of the $ O \left( n \log n \right) $, $ O \left( kn \log n \right) $, and $ O \left( kn \log n \right) + O \left( n \log n \right) $ algorithms to sort $ 2^{24} \left(x,y,z\right) $ tuples and construct a \emph{k}-d tree, plotted versus 1 to 16 threads. For this benchmark, single-threaded execution of the $ O \left( n \log n \right) $ \emph{k}-d tree-building phase of the $ O \left( kn \log n \right) + O \left( n \log n \right) $ algorithm was specified because dual-threaded execution is no faster. Multi-threaded execution via 1 to 16 threads was specified for the $ O \left( kn \log n \right) $ merge-sort phase of the $ O \left( kn \log n \right) + O \left( n \log n \right) $ algorithm, and for the $ O \left( kn \log n \right) $ and $ O \left( n \log n \right) $ algorithms.

Figure \ref{fig:time} demonstrates that the $ O \left( kn \log n \right) $ algorithm (red curve) is faster than the $ O \left( n \log n \right) $ algorithm (green curve), and that the $ O \left( kn \log n \right) $ and $ O \left( n \log n \right) $ algorithms are significantly faster than the $ O \left( kn \log n \right) + O \left( n \log n \right) $ algorithm (magenta curve), for 1 to 16 threads.

\begin{figure}[h]
\centering
\centerline{\includegraphics*[trim = {1.00in, 3.47in, 1.37In, 1.52In}, clip, width=\columnwidth]{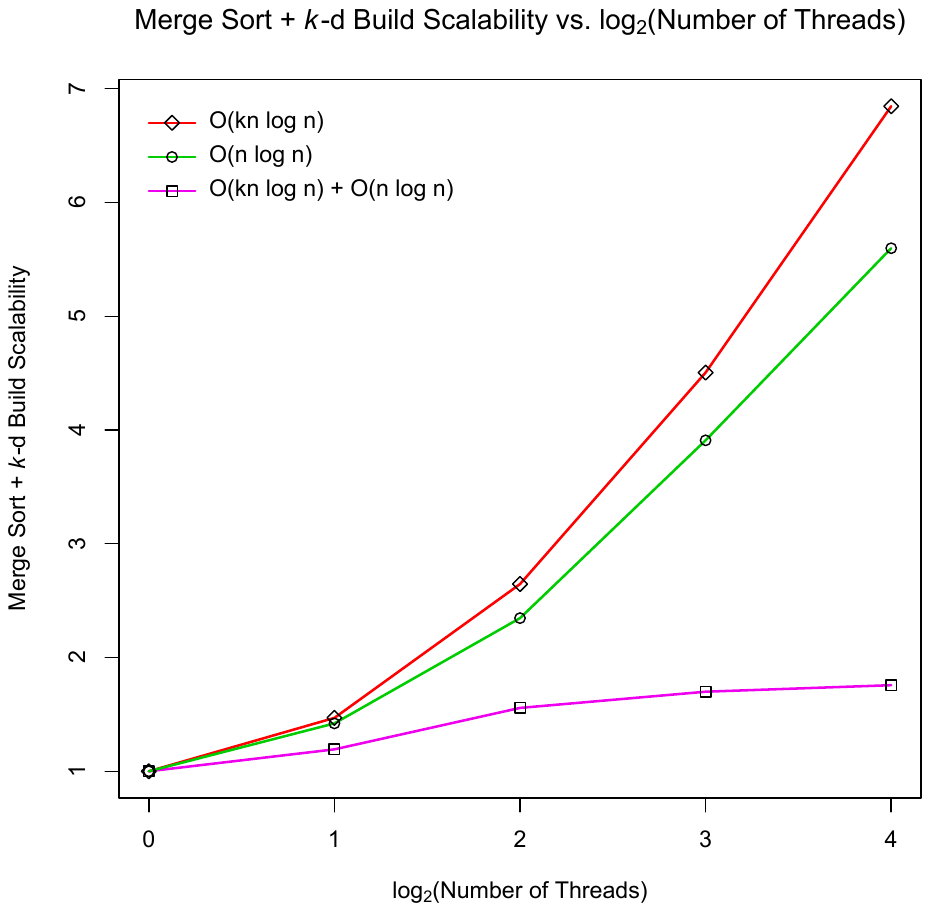}}
\caption{Scalability vs. $ \log_2 $(number of threads) for $ 2^{24} \left(x,y,z\right) $ tuples}
\label{fig:scale}
\end{figure}

Figure \ref{fig:scale} shows semi-log plots of the execution-time data from Figure \ref{fig:time} as \emph{scalability} instead of execution time. Scalability is calculated as $ t_1 / t_n $ where $ t_1 $ is the execution time for 1 thread, and $ t_n $ is the execution time for $ n \ge 1 $ threads.

This figure shows that the $ O \left( kn \log n \right) $ algorithm (red curve) scales better than the $ O \left( n \log n \right) $ algorithm (green curve), and that the $ O \left( kn \log n \right) $ and $ O \left( n \log n \right) $ algorithms scale significantly better than the $ O \left( kn \log n \right) + O \left( n \log n \right) $ algorithm (magenta curve), for 1 to 16 threads. For the $ O \left( kn \log n \right) + O \left( n \log n \right) $ algorithm, only the $ O \left( kn \log n \right) $ merge-sort phase is executed via multiple threads and hence only that phase scales.

\begin{figure}[h]
\centering
\centerline{\includegraphics*[trim = {1.00in, 3.47in, 1.37In, 1.52In}, clip, width=\columnwidth]{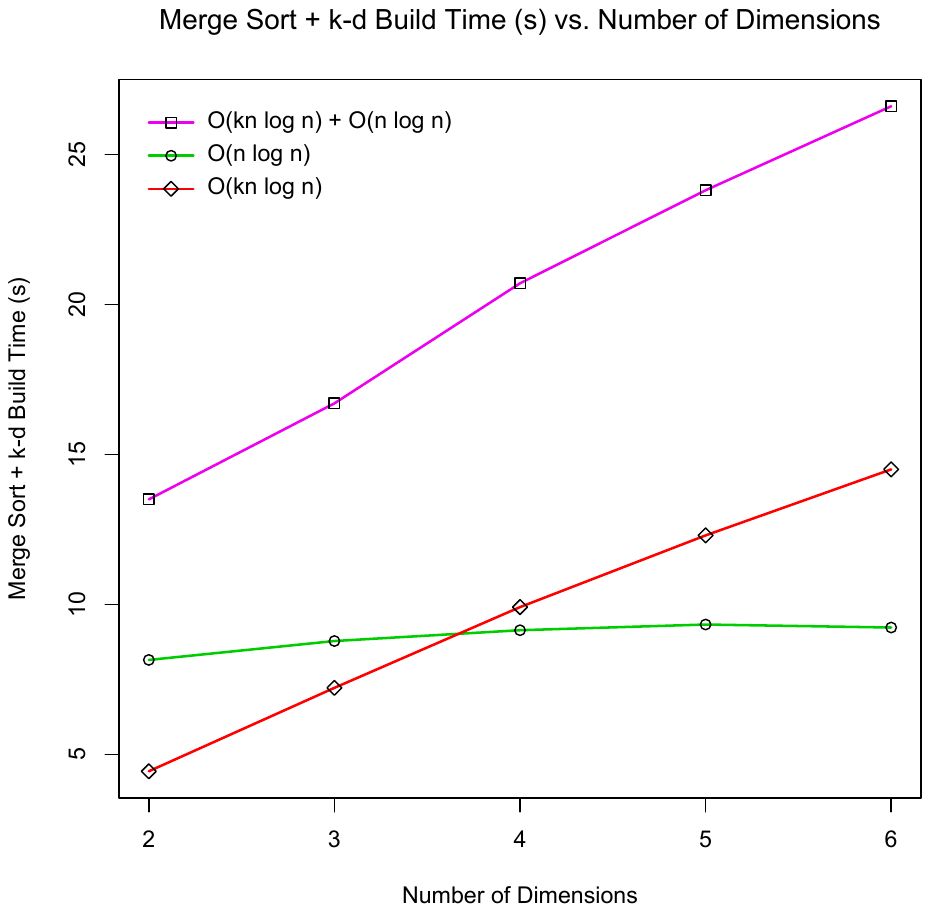}}
\caption{Execution time (s) vs. number of dimensions for  $2^{24} $ tuples and 4 threads}
\label{fig:dim4}
\end{figure}

Figure \ref{fig:dim4} plots the times in seconds for 4-threaded execution of the $ O \left( n \log n \right) $, $ O \left( kn \log n \right) $, and $ O \left( kn \log n \right) + O \left( n \log n \right) $ algorithms to sort $ 2^{24} $ \emph{k}-dimensional tuples and construct a \emph{k}-d tree, versus 2 to 6 dimensions \emph{k}.

This figure shows that the execution times of the $ O \left( kn \log n \right) $ algorithm (red curve) and the $ O \left( kn \log n \right) + O \left( n \log n \right) $ algorithm (magenta curve) increase linearly with \emph{k}, whereas the execution time of the $ O \left( n \log n \right) $ algorithm (green curve) does not increase with \emph{k}. These results are consistent with the computational complexities of the algorithms.

\newpage

\begin{figure}[h]
\centering
\centerline{\includegraphics*[trim = {1.00in, 3.47in, 1.37In, 1.52In}, clip, width=\columnwidth]{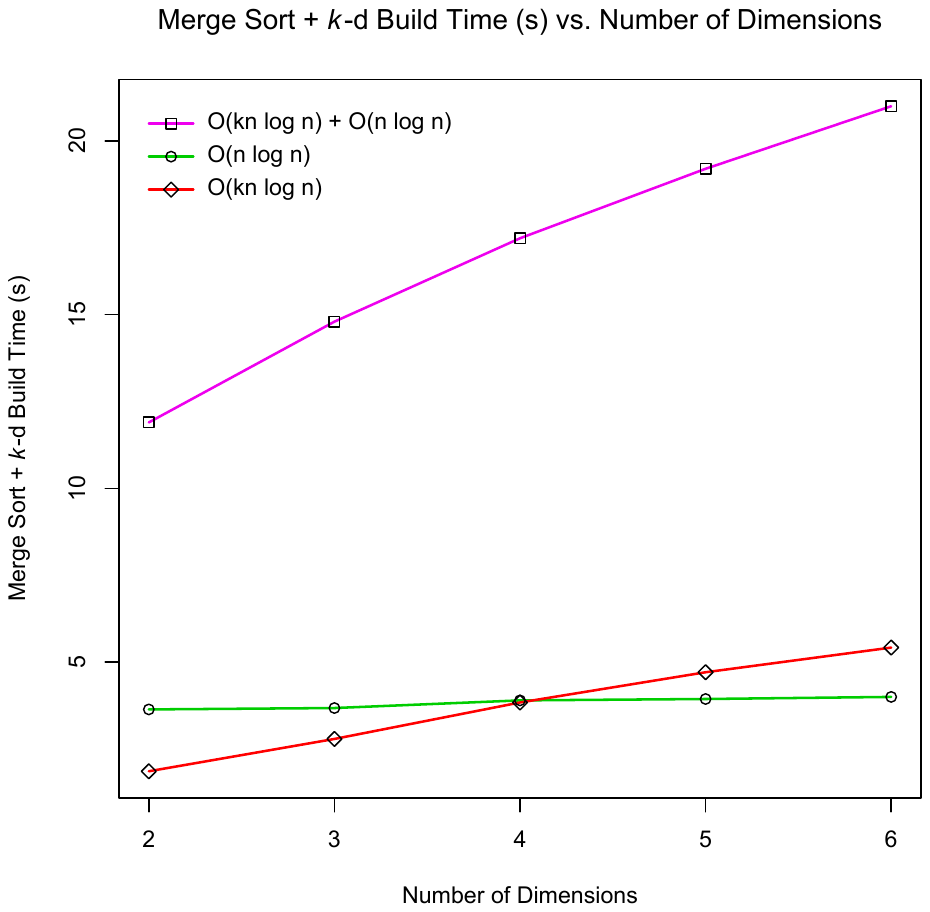}}
\caption{Execution time (s) vs. number of dimensions for  $2^{24} $ tuples and 16 threads}
\label{fig:dim16}
\end{figure}

Figure \ref{fig:dim16} plots the times in seconds for 16-threaded execution of the $ O \left( n \log n \right) $, $ O \left( kn \log n \right) $, and $ O \left( kn \log n \right) + O \left( n \log n \right) $ algorithms to sort $ 2^{24} $ \emph{k}-dimensional tuples and construct a \emph{k}-d tree, versus 2 to 6 dimensions \emph{k}.

This figure shows that the red curve of the $ O \left( kn \log n \right) $ algorithm crosses the green curve of the $ O \left( n \log n \right) $ algorithm at a larger value of $ k \left(=  4\right) $ than in Figure \ref{fig:dim4}, and is consistent with previously reported results \cite{Brown2015}. Crossing at a larger value of $ k \left(= 4\right) $ is expected, because Figure \ref{fig:scale} demonstrates that the increased scalability of the $ O \left( kn \log n \right) $ algorithm relative to the $ O \left( n \log n \right) $ algorithm is greater for 16 threads than for 4 threads.

\newpage

\section{Conclusions}
\label{Conclusions}

For building a \emph{k}-d tree that comprises $ 2^{24} $ tuples, the $ O \left( n \log n \right) $ algorithm is slower than the $ O \left( kn \log n \right) $ algorithm for three or fewer dimensions \emph{k}, but faster for five dimensions or more. The $ O \left( n \log n \right) $ algorithm does not scale as well as the $ O \left( kn \log n \right) $ algorithm. Future research could be directed improving the scalability of the median of medians algorithm to improve the scalability of the $ O \left( n \log n \right) $ algorithm.

Cao et al. have proposed an innovative algorithm for building a \emph{k}-d tree that obviates the need to partition pre-sorted index arrays, but instead registers the partition of those arrays in a data structure. This new algorithm is not amenable to execution by more than two threads, and its dual-threaded performance is no faster than its single-threaded performance. The algorithm is slow compared to the $ O \left( n \log n \right) $ and $ O \left( kn \log n \right) $ algorithms due to cache contention. Future research could be directed at creation of a different data structure for this algorithm to facilitate multi-threaded execution and minimize cache contention.

\section*{Supplemental Materials}

Included with this manuscript are C++ implementations of $ O \left( n \log n \right) $, $ O \left( kn \log n \right) $, and $ O \left( kn \log n \right) + O \left( n \log n \right) $ algorithms that build a \emph{k}-d tree, and $ O \left( n \log n \right) $ and $ O \left( kn \log n \right) $ algorithms that build a \emph{k}-d tree-based key-to-value map.

\section*{Author Contact Information}

\href{https://www.linkedin.com/in/russellabrown/}{https://www.linkedin.com/in/russellabrown/}


\newpage

\section*{References}

\small
\bibliographystyle{jcgt}
\bibliography{paper}

@ARTICLE{Bentley,
  author = {J.L. Bentley},
  title = {Multidimensional binary search trees used for associative searching},
  journal = {Communications of the ACM},
  volume = {18},
  issue = {9},
  pages = "509-517",
  year = 1975,
  url = {https://dl.acm.org/toc/cacm/1975/18/9},
  doi = {10.1145/361002.361007}
}

@ARTICLE{Friedman,
  author = {J.H. Friedman and J.L. Bentley and R.A. Finkel},
  title = {An algorithm for finding best matches in logarithmic expected time},
  journal = {ACM Transactions on Mathematical Software},
  volume = {3},
  issue = {3},
  pages = "209-226",
  year = 1977,
  url = {http://dl.acm.org/citation.cfm?id=355745},
  doi = {10.1145/355744.355745}
}

@ARTICLE{Blum,
  author = {M. Blum and R.W. Floyd and V. Pratt and R.L. Rivest and R.E. Tarjan},
  title = {Time bounds for selection},
  journal = {Journal of Computer and System Sciences},
  volume = {4},
  pages = "448-461",
  year = 1973,
  url = {http://people.csail.mit.edu/rivest/BlumFloydPrattRivestTarjan-TimeBoundsForSelection.pdf},
  doi = {10.1016/S0022-0000(73)80033-9}
}

@INCOLLECTION{Stepanov,
   author = {A. Stepanov and P. McJones},
   title = {Linear Orderings},
   booktitle = {Elements of Programming}, 
   edition = {First},
   pages = "49-63", 
   year = 2009, 
   publisher = {Addison-Wesley},
   address = {New York},
   url = {https://elementsofprogramming.com/}
}

@INCOLLECTION{Procopiuc,
   author = {O. Procopiuc and P.K. Agarwal and L. Arge and J.S. Vittner},
   title = {Bkd-tree: {A} dynamic scalable kd-tree},
   booktitle = {Lecture Notes in Computer Science (LNCS), Advances in Spatial and Temporal Databases}, 
   editor = {T. Hadzilacos and Y. Manolopoulos and J. Roddick and Y. Theodoridis},
   volume = {2750},
   pages = "46-65", 
   year = 2003, 
   publisher = {Springer-Verlag},
   address = {Berlin},
   url = {https://link.springer.com/chapter/10.1007/978-3-540-45072-6_4},
   doi = {10.1007/978-3-540-45072-6\_4}
}

@ARTICLE{Brown2015,
  author = {R. A. Brown},
  title = {Building a balanced \emph{k}-d tree in \emph{{O}}(\emph{kn} log \emph{n}) time},
  journal = {Journal of Computer and Graphics Techniques (JCGT)},
  volume = {7},
  pages = "50-68",
  year = 2015,
  url = {http://jcgt.org/published/004/01/03/}
}

@ARTICLE{Brown2025,
  author = {R. A. Brown and J. A. Robinson},
  title = {Appendix to: {B}uilding a balanced \emph{k}-d tree in \emph{{O}}(\emph{kn} log \emph{n}) time},
  journal = {arXiv},
  volume = {1410.5420},
  pages = "1-11",
  year = 2025,
  url = {https://arxiv.org/abs/1410.5420},
  doi = {10.48550/arXiv.1410.5420}
}

@ARTICLE{Cao,
   author = {Y. Cao and H. Wang and W. Zhao and B. Duan and X. Zhang},
   title = {A new method to construct the {KD} tree based on presorted results},  
   year = {2020},
   journal = {Complexity},
   volume = {2020},
   number = {8883945},
   pages = "1-7",
   url = {https://onlinelibrary.wiley.com/doi/10.1155/2020/8883945},
   doi = {10.1155/2020/8883945}
}

@ARTICLE{Matsumoto,
   author = {M. Matsumoto and T. Nishimura},
   title = {Mersenne Twister: A 623-Dimensionally Equidistributed Uniform Pseudo-Random Number Generator},  
   year = {1998},
   journal = {{ACM} Transactions on Modeling and Computer Simulation},
   volume = {8},
   pages = "3-30",
   url = {https://dl.acm.org/doi/10.1145/272991.272995},
   doi = {10.1145/272991.272995}
}

@INCOLLECTION{Chandra,
   author = {R. Chandra and L. Dagum and D. Kohr and D. Maydan and J. McDonald and R. Menon},
   title = {False Sharing},
   booktitle = {Parallel Programming in OpenMP}, 
   edition = {First},
   pages = "189-191", 
   year = 2001, 
   publisher = {Academic Press},
   address = {London}
}

\end{document}